\documentclass[8pt,a4paper]{article}
\newcommand{\dbar}{d\hspace*{-0.08em}\bar{}\hspace*{01.em}}
\usepackage[utf8]{inputenc}
\usepackage{amsmath}
\usepackage{amsfonts}
\usepackage{amssymb}
\usepackage{graphicx}
\usepackage{xcolor}
\usepackage{url}
\usepackage{braket}
\usepackage{tikz}
\usepackage[left=3cm, right=3cm, top=2cm, bottom=2cm]{geometry}
\usepackage{rotating}
\usepackage{lscape}
\usepackage{hyperref}
\usepackage{lscape}
\usepackage{wrapfig}
\usepackage{float}
\usepackage{pdfpages}

\newcommand{\rl}[1]{\left(#1\right)}

\allowdisplaybreaks

\def\be{\begin{equation}}
\def\ee{\end{equation}}
\def\ba{\begin{array}{l}}
	\def\ea{\end{array}}
\def\bea{\begin{eqnarray}}
\def\eea{\end{eqnarray}}
\def\beas{\begin{eqnarray*}}
	\def\eeas{\end{eqnarray*}}

\begin{document}
	\baselineskip 24pt
	\begin{center}
		
	{\large \textbf{QUANTUM THERMODYNAMICS OF SMALL SYSTEMS:\\ THE ANYONIC 
OTTO ENGINE}\par}
		
	\end{center}
	
	\vskip .5cm
	\medskip
	
	\vspace*{4.0ex}
	
	\baselineskip=18pt
	
\begin{center}
\large
H S Mani$^{(a)}$\footnote {hsmani@cmi.ac.in}, 
Ramadas N$^{(b)}$\footnote{ic37306@imail.iitm.ac.in}, 
and V V Sreedhar$^{(a)}$\footnote{sreedhar@cmi.ac.in}

\end{center}
	\vspace*{4.0ex}

	\centerline{\it $^{(a)}$ Chennai Mathematical Institute,  SIPCOT IT Park, Siruseri, Chennai, 603103 India} 
	
	\centerline{\it $^{(b)}$ Department of Physics, Indian Institute of
Technology Madras, Chennai, 600036 India} 
	
	\vspace*{1.0ex}

	\vspace*{1.0ex}

	\vspace*{5.0ex}

	\begin{abstract}
Recent advances in applying thermodynamic ideas to quantum systems have raised 
the novel prospect of using non-thermal, non-classical sources of energy, of 
purely quantum origin, like quantum statistics, to extract mechanical work in 
macroscopic quantum systems like Bose-Einstein condensates. On the other 
hand, thermodynamic ideas have also been applied to small systems like single
molecules and quantum dots. In this paper we study the quantum thermodynamics 
of small systems of anyons, with specific emphasis on the quantum Otto 
engine which uses, as its working medium, just one or two anyons. Formulae 
are derived for the efficiency of the Otto engine as a function of the 
statistics parameter. 
		
	\end{abstract}
\section{Introduction} 

Thermodynamics is an empirical and phenomenological description of 
matter at the macroscopic level, where the number of particles in the 
system is of the order of the Avogadro number \cite{callen}. It is of 
academic interest to stretch the thermodynamic line of thinking to small 
systems in order to probe the limits of applicability of concepts like 
temperature and entropy \cite{feshbach}. In the last few decades, several 
small systems, like single molecules and quantum dots, have been studied 
extensively from a thermodynamic point of view \cite{terrell}. These 
studies are made possible by bringing the ideas of quantum mechanics and 
thermodynamics under one umbrella, with the obvious name of quantum 
thermodynamics, which allows us to push the frontiers of thermodynamics 
to the microscopic level.         

When two disparate approaches to physical problems face off, as in the above 
case, surprises are to be expected. Classical thermodynamic engines like the 
Otto engine, studied and used for over a century, convert thermal 
energy to work. On the other hand, quantum thermodynamic engines afford us an 
opportunity to harness non-classical, non-thermal sources of energy, arising 
out of quantum statistics, to do mechanical work. 

A simple back-of-the-envelope calculation reveals that for a harmonically 
trapped quantum bose gas of $N$ particles, the energy at zero temperature 
is $E^B = N\hbar\omega/2$ since all the particles occupy the ground state, 
whereas, for a fermi gas, for which all levels upto the Fermi energy 
$E_{Fermi}= \hbar\omega(2N-1)/2$, are occupied, it is $E^F = \hbar\omega 
N^2/2$. The difference in these energies, $E_P = E^F - E^B = \hbar\omega 
N(N-1)/2$, with its origin in the exclusion principle, and hence called the 
Pauli energy, can, in principle, be tapped by a quantum engine, and can be 
very large for large values of $N$, {\it i.e.} for macroscopic systems. This 
energy is non-classical, and purely quantum mechanical in origin, derived as it 
is from the quantum statistical population distribution functions of 
indistinguishable particles. 

A recent paper by Koch {\it et al} \cite{koch} reports an experimental 
realization of the above idea by constructing a many-body quantum engine, 
fittingly called the Pauli engine, with harmonically trapped  $~^6$Li atoms 
close to a magnetic Feshbach resonance. Like the classical Otto engine, the 
Pauli engine consists of four strokes {\it viz.} compression, fermionization, 
expansion, and bosonization. The change in quantum statistics of the gas is 
accomplished by tuning the magnetic field to drive the quantum gas back 
and forth between a Bose-Einstein condensate and a unitary Fermi gas, through
the well-known phenomenon of BEC-BCS crossover \cite{leggett}.

In this, the first of two papers on the topic, we study the quantum 
thermodynamics of small systems, consisting of one or two anyons, to be 
precise, whose quantum statistics can be made to smoothly interpolate between 
the bosonic and fermionic limits, and construct an Otto engine which converts 
a change of quantum statistics to mechanical work in one dimension.  

In section {\bf 2}, we briefly review the formalism of quantum thermodynamics,
with particular emphasis on the difference between a classical and quantum 
Otto engine. In section {\bf 3}, we define the basic model of a quantum Otto
engine based on anyons. In section {\bf 4}, we advance a charged particle 
constrained to move on a ring threaded by a magnetic flux, as a model of a 
one-dimensional anyon. In section {\bf 5}, we derive analogous results for 
two anyons on a ring in the Calogero-Sutherland model. For both the models we 
set up the quantum Otto engine and calculate its efficiency. We conclude with 
a few closing remarks in section {\bf 6}.               

\section{Quantum Thermodynamics}

The main idea of quantum thermodynamics is to identify the non-classical 
equivalents of thermodynamic concepts like internal energy, heat, and work 
in a quantum system \cite{qtd}. 

Let $\rho$ be a density operator that describes a quantum system coupled 
to a thermal environment. Let $H(\lambda )$ be the system Hamiltonian, and 
$\lambda$ be a control parameter. The internal energy is defined by 
\begin{equation} 
E =~ <H>~ =~ {\hbox{tr}}\{\rho H\}
\end{equation} 
where, in the weak coupling limit,  $\rho$ is the  equilibrium state of 
the system which we take to be the Gibbs' state
\begin{equation} 
\rho = {1\over Z} {\hbox{exp}}(-\beta H) 
\end{equation} 
where $Z = {\hbox{tr}}~{\hbox{exp}}(-\beta H)$ is the partition function,
with $\beta = 1/k_BT$ as usual. As is well-known, the entropy for a 
Gibbs' state is given by $S = - k_B~{\hbox{tr}}~\{\rho{\hbox{ln}}\rho\}$. 

The change in internal energy can be partitioned into two pieces {\it viz.} 
\begin{equation} 
dE = {\hbox{tr}}\{d\rho~ H\} + {\hbox{tr}}\{\rho~dH\} = \dbar Q + \dbar W
\end{equation} 
The first term represents a change in entropy while the second term 
represents a change in the Hamiltonian, the $\dbar$ indicating that neither
of these changes is exact. In complete analogy with classical 
thermodynamics, we conclude that the work done corresponds to a displacement in 
the energy levels, and heat corresponds to a change in the probability 
distribution that populates the energy levels. 

It is now straightforward to define quantum analogues of isothermal, 
isobaric, isochoric, and adiabatic processes, and hence the various 
quantum analogues of the classical thermodynamic engines.

The Otto engine, for example, consists of four strokes: two adiabats 
and two isochores. By definition, an adiabatic process is one in 
which no heat transfer takes place between the system and the 
environment and this corresponds to a change in the energy eigenvalues
while keeping the populations, and hence the von Neumann entropy, unchanged. 
An isochoric process, on the other hand, keeps the energy eigenvalues 
fixed while allowing for changes in the populations of these levels.   

To conclude this brief survey of quantum thermodynamics, we need to 
mention a few subtle points in which the classical and quantum 
versions of thermodynamics differ. 

A classical adiabatic process is characterised by complete thermal insulation 
because of which no heat can be exchanged with the environment. A quantum 
adiabat on the other hand follows the adiabatic theorem in which the relevant 
eigenstate is dragged through the process. It is not possible to maintain a 
quantum adiabat for a long time because of decoherence. Thus, the time-scale of 
the adiabat should be less than the decoherence time-scale. 

Unlike classical thermodynamical engines which are reversible, and are in 
instantaneous equilibrium through out, in quantum thermodynamic engines, 
finite-time adiabats drive the system out of equilibrium, and a relaxation 
process is necessary for a new equilibrium state to be reached through 
thermalization with a bath.  

Quantum Otto engines based on qubits, three-level systems, harmonic 
oscillators, and statistical anyons \cite{myers}  have been extensively 
studied. In this paper we study the quantum Otto engine with a working medium
being a very small number of one-dimensional anyons -- particles which 
intrinsically have any quantum statistics, and which can smoothly interpolate 
between the bosonic and fermionic limits. 

\section{An Anyonic Quantum Otto Process}

As is well-known, the spin and statistics theorem in quantum theory allows 
for two types of particles: a) bosons, which have integer spin, have
wavefunctions which transform under the symmetric representation of the 
permutation group, and follow the Bose-Einstein statistical distribution, 
and b) fermions, which have half-odd integer spin,  have wavefunctions
which transform under the alternating representation of the permutation 
group, and follow the Fermi-Dirac distribution \cite{pauli}. 

In low dimensions, spin and statistics can take arbitrary values, and particles with such properties are called anyons. The underlying topological reasons for 
these more general possibilities have been extensively studied in two 
dimensions \cite{leinaas}. On a real line, an exchange of two 
indistinguishable particles requires us to take one particle through the 
other, and thus gets inextricably linked with interaction. This very fact 
allows us to define exchange statistics. In the next couple of sections, we 
consider two such models. The first is that of a charged particle constrained 
to move on a circular ring threaded by a magnetic flux which influences the 
periodicity properties of the particle's wavefunction \cite{wilczek}. The 
second realises anyonic statistics through an interaction between two 
particles as described by the Calogero-Sutherland Hamiltonian \cite{sutherland}. 
\section{Charged Particle On A Ring Threaded By A Magnetic Flux Tube}
In this example, we have an infinitely long solenoid of cross-sectional 
area $A$, carrying a magnetic field $ (0,0,B) $. The magnetic flux is 
$\Phi =BA$. 
\begin{figure}
\centering
\includegraphics[scale=0.25]{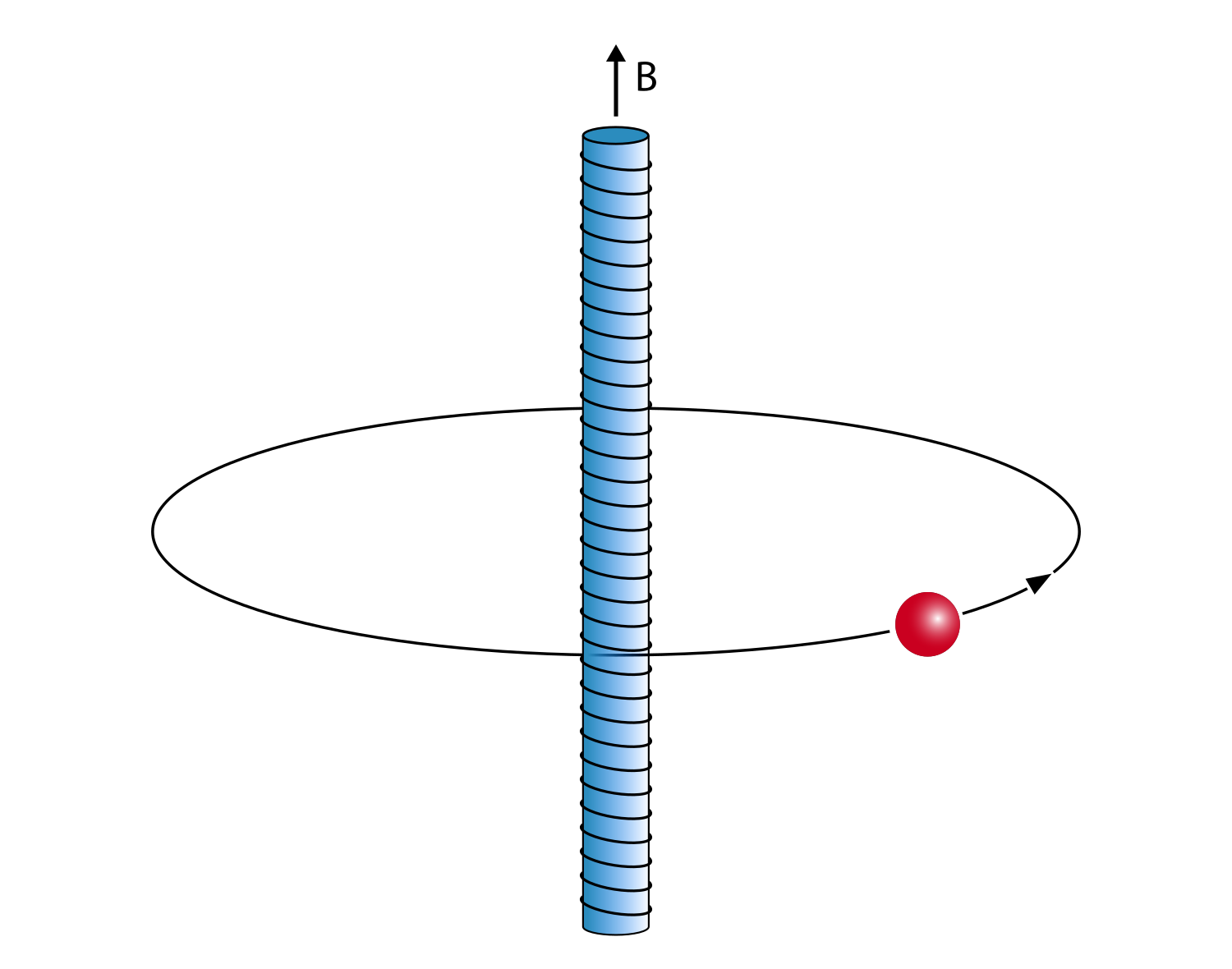}
\caption{Charge Circling A Magnetic Flux Tube}
\end{figure} 

Although it doesn't make sense to talk about statistics of individual 
particles, this may be considered as a toy model of an anyon on a ring of 
radius $a$. To verify this statement all we have to do is to consider two 
particles on the ring and exchange their positions. 
\begin{figure}[h]
\centering
\includegraphics[scale=0.25]{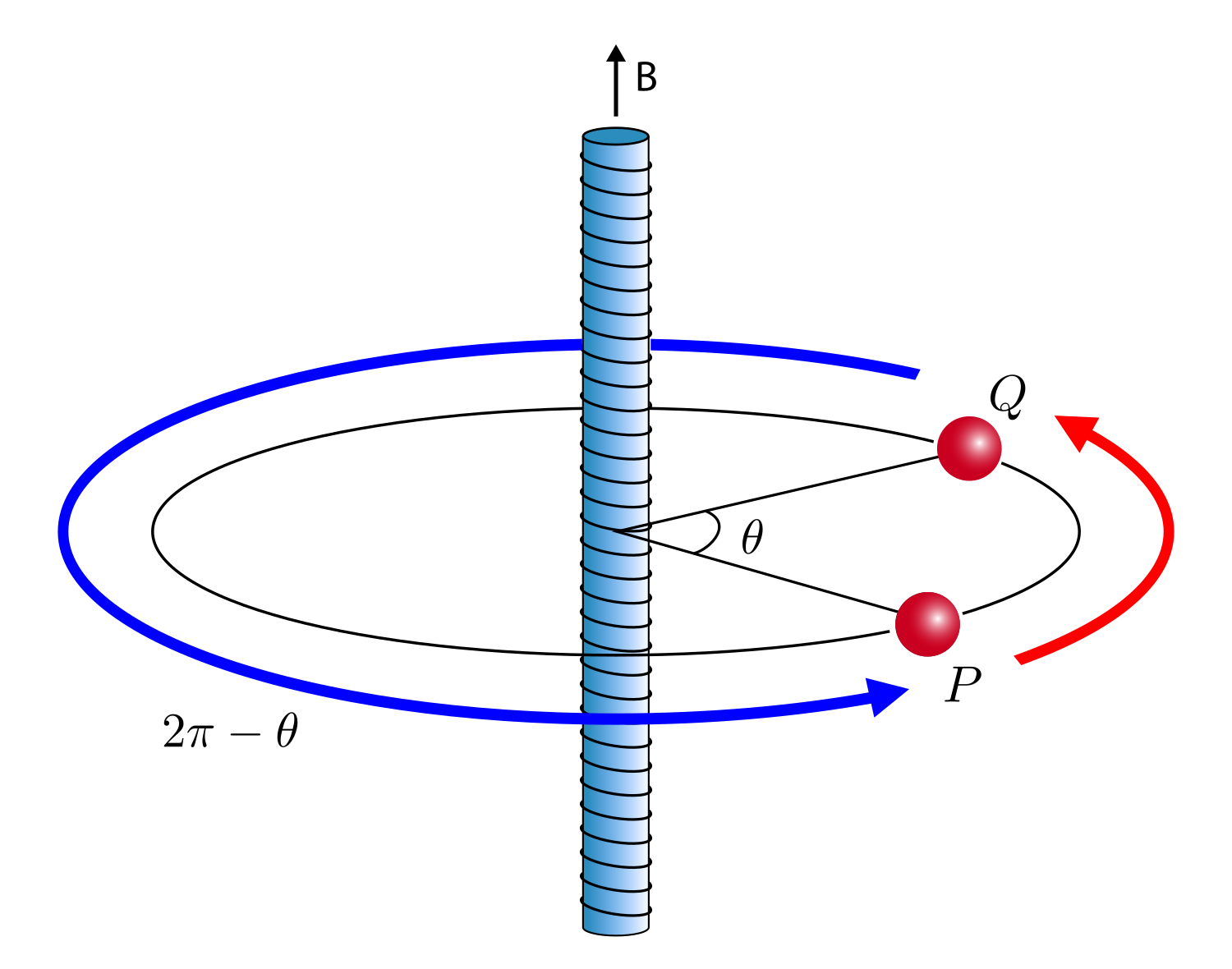}
\caption{Two Anyons On A Ring}
\end{figure}

As already mentioned, it is not possible to exchange particle positions in one
dimensional space ({\bf $R^1$}) without taking them through each other. 
However, as can be seen from the figure, this problem can be bypassed for two 
particles on a ring. The two-particle wavefunction thus picks up an 
Aharonov-Bohm phase ${\hbox{exp}}~ [2i\pi\Phi]$ under an exchange, which 
may be interpreted as the phase factor acquired in exchanging anyons.    

The vector potential has only the azimuthal component
\begin{align}
\begin{aligned}
A_{\phi} = \frac{\Phi}{2\pi r}
\end{aligned}
\end{align}
The Hamiltonian of a charged particle $q$ on the ring is 
\begin{align}
\begin{aligned}
H = \frac{1}{2m} \rl{p_\phi - q A_{\phi}}^2 = \frac{1}{2ma^2} \rl{ -i\hbar 
\frac{\partial }{\partial \phi} - \frac{q\Phi}{2\pi} }^2
\end{aligned}
\end{align}
The normalised energy eigenstates are 
\begin{align}
\begin{aligned}
\psi_n(\phi)= \frac{1}{\sqrt{2\pi} } e^{in\phi}, ~~n \in \mathbb{Z}
\end{aligned}
\end{align}
with energy eigenvalues
\begin{align}
\begin{aligned}
E_n = \frac{\hbar^2}{2ma^2}\rl{n- \alpha}^2,~~~\alpha = \frac{q\Phi}{2\pi\hbar}
\end{aligned}
\end{align}

\subsection{Quantum Otto Engine}
A schematic diagram of the quantum Otto engine is given below. 
\begin{figure}[h]
\centering
\includegraphics[scale=0.5]{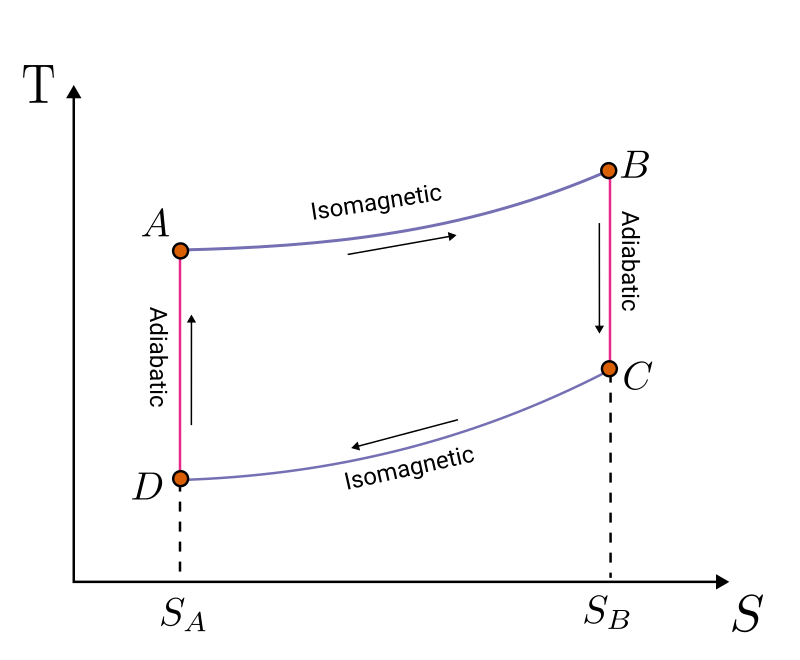}
\caption{The Quantum Otto Engine}
\end{figure}

The four strokes that constitute the quantum Otto engine are as follows:
In the first step, as we move from $A$ to $B$, the system changes its
temperature from $T_l$ to $T_h$. This is achieved by bringing the system 
in contact with an infinite bath at each infinitesimal temperature step as
$T_l\rightarrow T_l+\Delta T \rightarrow T_l + 2\Delta T, \cdots 
\rightarrow T_l + (N-1)\Delta T \rightarrow T_l+N\Delta T = T_h$. 
As $\Delta T \rightarrow 0, ~N\rightarrow\infty$, such that 
$N\Delta T = T_h - T_l$, we get a reversible path. This is because the 
entropy change of the reservoir and the system is zero at each stage. 
Similar arguments hold for the path $C$ to $D$.

$A \to B $ and $ C\to D $ are isomagnetic processes, thus called because 
no work is done along these paths. Recall that changes in energy levels 
(quantum work) are effected by changes in the magnetic field. The 
strength of the magnetic field on $ A \to B $ is chosen to be $B_h$, and 
correspondingly, the energy is $ E_n^h =  \frac{\hbar^2}{2ma^2} 
\rl{n- \alpha_h}^2 $, where $\alpha_h = \pi a^2 B_h$.  Similarly, from $ C 
\to D $ the energy levels are  $ E_n^l =  \frac{\hbar^2}{2ma^2} 
\rl{n- \alpha_l}^2 $ and the corresponding magnetic field is $ B_l$. The change 
in the magnetic field along the adiabats produces a current which can be 
translated to mechanical work. 

Alternatively, we can keep the magnetic field constant and change the radius 
of the ring, {\it i.e.} along $AB$ and $CD$, the energy levels are given by 
$E^h_n = {\hbar^2\over ma_1^2} (n-\pi a_1^2B)$ and 
$E^l_n = {\hbar^2\over ma_2^2} (n-\pi a_2^2B)$ respectively. 

If $a_2>a_1$, $E_n$ decreases, {\it i.e.} $E_n^h > E_n^l$, but since the 
occupation probabilities $P_n$ remain the same in the adiabatic processes 
$ B \to C $ and $ D \to A $, work is done by the system as we go from $B$ 
to $C$, and on the system as we go from $D$ to $A$. In both cases, the 
entropy remains the same. Note that only the states $B$ and $D$ are in 
thermal equilibrium, but not $A$ and $C$.\footnote{It should be mentioned 
that for an adiabatic process, the temperature of systems with more than 
two levels is in general not defined. For systems with more than two levels, 
one needs to allow for effects of relaxation, as already mentioned. We can 
ignore this complication if we restrict ourselves to sufficiently low 
temperature, and hence, to the lowest two levels \cite{selcuk}.}

The efficiency of the quantum Otto cycle is give by
\begin{align}
\begin{aligned}
\eta_{QOE}=  \frac{W_{out}}{Q_{in}} =1-\frac{\sum_n E^l_n(P_n(B) -P_n(A))  }{ \sum_m E^h_m (P_m(B) -P_m(A))   }
\end{aligned}
\end{align}\\
where $ P_n(X) = \frac{e^{-\beta_X E_n(X)}}{Z(X)} $.
It is easy to check that each term in the summand in the numerator is 
less than the corresponding term in the denominator since $P_n(B) - P_n(A) >0$
as we move from lower to higher temperature, and $E_n^l < E_n^h$ for $a_1<a_2$
consistent with $1>\eta>0$.  Using the expressions  
\begin{align}
\begin{aligned}
P_n(B) = P_n(C)  &= \frac{ e^{-\beta_h E^h_n}} {\sum_n e^{-\beta_h E^h_n} } \\
P_n(A)=P_n(D) &= \frac{ e^{-\beta_l E^l_n}} {\sum_n e^{-\beta_l E^l_n} } \\
\end{aligned}
\end{align}
we write
\begin{align}
\begin{aligned}
\eta &= 1- \frac{\rl{ \frac{\sum_n E^l_n  e^{-\beta_h E^h_n}} {\sum_{n_1} e^{-\beta_h E^h_{n_1}} }  - \frac{ \sum_n E^l_n e^{-\beta_l E^l_{n}}} {\sum_{n_2} e^{-\beta_l E^l_{n_2}} }   }  }{\rl{ \frac{ \sum_n E^h_m  e^{-\beta_h E^h_m}} {\sum_{m_1} e^{-\beta_h E^h_{m_1}} }  - \frac{ \sum_n E^h_m  e^{-\beta_l E^l_{m}}} {\sum_{m_2} e^{-\beta_l E^l_{m_2}} }   } } 
\end{aligned}
\end{align}
The sums appearing in the above equation can be calculated in a straighforward
manner, and give the following analytic expression for the efficiency of the 
anyonic quantum Otto engine: 
\begin{align}
\begin{aligned}
  \eta 
  & = 1-\frac{ \frac{\Upsilon (l,h)}{Z_h} -\frac{\Upsilon (l,l)}{Z_l}  }{ \frac{\Upsilon (h,h)}{Z_h} -\frac{\Upsilon (h,l)}{Z_l}  }
\end{aligned}
\end{align}
where 
\begin{align}
\begin{aligned}
\Upsilon(k,j) &=  \sum_{n=-\infty}^{\infty} E^k_n e^{ -\beta_j E^j_n},~~~~
j,k = h,l\\
& =   \sum_{n=-\infty}^{\infty} \frac{\hbar^2}{2ma^2} \rl{n- \alpha_k}^2  
e^{ -\beta_j \frac{\hbar^2}{2ma^2} \rl{n- \alpha_j}^2 } \\
& =  \frac{\hbar^2}{2ma^2} \bigg( c^2 e^{-\lambda \gamma^2}\vartheta _3 
\rl{e^{2 \lambda \gamma},e^{-\lambda}}+ e^{-\lambda \gamma^2} 
\frac{(c\gamma)}{ \lambda}  \frac{\partial }{\partial \gamma} 
\vartheta _3 \rl{e^{2 \lambda \gamma},e^{-\lambda}}\\
& -e^{-\lambda \gamma^2}  \frac{\partial }{\partial \lambda}\vartheta _3 
\rl{e^{2 \lambda \gamma},e^{-\lambda}} \bigg) \bigg|_{c=\alpha_k, \lambda = 
\beta_j \frac{\hbar^2}{2ma^2}, \gamma = \alpha_j }
\end{aligned}
\end{align}
and the partition function is given by  
\begin{align}
\begin{aligned}
Z_j &= \sum_{n=-\infty}^{\infty} e^{ -\beta_j E^j_n}  , ~~j=h,l \\
&= \sum_{n=-\infty}^{\infty}e^{-\beta_j\frac{\hbar^2}{2ma^2} 
\rl{n- \alpha_j}^2 }  \\
& =  e^{ -\beta_j \frac{\hbar^2}{2ma^2} \alpha_j^2 } \vartheta _3 
\rl{\frac{\alpha_j \beta_j \hbar^2}{2ma^2},e^{-\beta_j \frac{\hbar^2}{2ma^2} }} 
\end{aligned}
\end{align}
The Jacobi theta function $\vartheta_3 \rl{x,q}$ in terms of which the above
expressions are written, is defined by
\begin{align}
\begin{aligned}
\vartheta _3 \rl{x,q} &= \sum_{n=-\infty}^{\infty} q^{n^2} x^n
\end{aligned}
\end{align}
The detailed calculations of the above results are relegated to the Appendix.

\section{Two Anyons On A One-Dimensional Ring}

Consider a system of two particles on a ring of finite circumference ($2\pi L$)
with periodic boundary conditions \cite{sutherland}. The Hamiltonian of the 
system is 
\begin{align}
\begin{aligned}
H = -\frac{\hbar^2}{2m}\sum_j \frac{\partial^2}{\partial x_j^2} + \frac{\pi^2 \alpha  
(\alpha-1)}{L^2}\sum_{j<k} \frac{1}{\sin^2 \rl{\frac{\pi (x_j-x_k)}{L^2}}}
\end{aligned}
\end{align}
In this case, the magnetic field of the previous section is replaced by an 
interaction between the two particles, with the strength of the interaction 
being directly related to the quantum statistics of the two particles. 

Setting $\hbar = m = 1$ to avoid clutter, the energy levels of the two-particle 
system are
\begin{align}
\begin{aligned}
E_{n_1,n_2} (L) = \frac{\pi^2 \alpha^2}{L^2} +\frac{2\pi^2 }{L^2} \rl{ n_1^2+n_2^2 +\alpha (n_1-n_2)}
\end{aligned}
\end{align}
where $ n_1,n_2 $ are integers, $ n_1 \leq n_2 $. The corresponding energy 
eigenstates are:
\begin{equation} 
\psi (\theta )= \Phi (\theta )\Delta^{\alpha}(\theta )
\end{equation} 
with $\Phi$ being a symmetric polynomial in the variables $z_i=e^{i\theta_j}$ 
and $z_j^{-1}$, $\theta$ being related to the coordinates by the equation 
$\theta_j = 2\pi x_j/L$, and the Jastrow factor being  
\begin{equation}
\Delta (\theta ) = \prod_{i<j} {\hbox{sin}}({\theta_i -\theta_j\over 2})
\end{equation}
The antisymmetry of $\Delta$ implies, in particular, that $\alpha = 0$ and 
$\alpha = 1$ correspond to bosons and fermions respectively. For other 
intermediate values, the particles have anyonic statistics. 

\subsection{The Quantum Otto Engine}

The two volumes can be chosen as $ V_1 = L_1 $ and $ V_2 = L_2 $.  The inverse 
temperature of the hot reservoir is $ \beta_h $ and that of the cold 
reservoir is $ \beta_l $. Energy levels are labeled by $ (n_1,n_2) $ where 
$ n_1 \leq n_2 $.
\begin{figure}[h]
\centering
\includegraphics[scale=0.5]{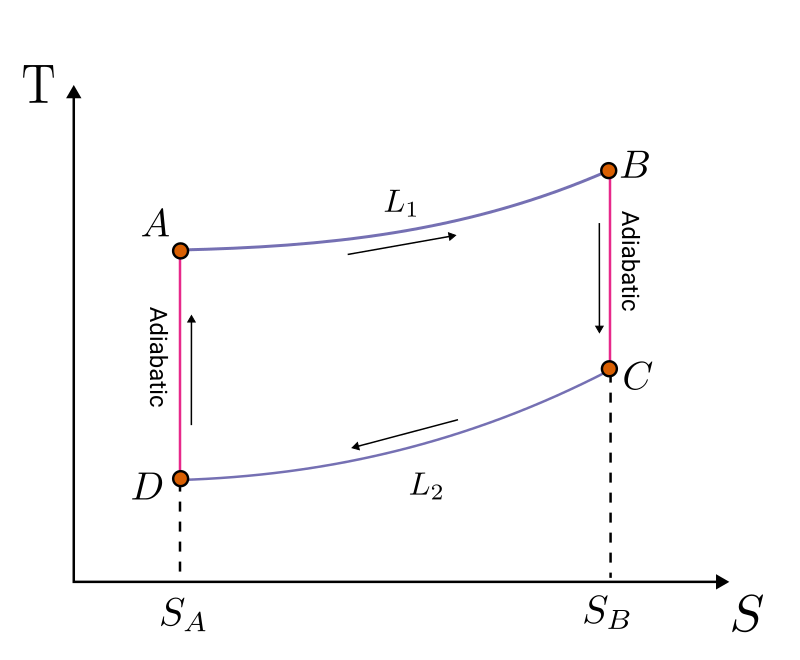}
\caption{The Quantum Otto Engine: variable volume, fixed coupling}
\end{figure}
We have 
\begin{align}
\begin{aligned}
E_{n_1,n_2} (L_1) &= \frac{\pi^2 \alpha^2}{L_1^2} +\frac{2\pi^2 }{L_1^2} \rl{ n_1^2+n_2^2 +\alpha (n_1-n_2)} \\
E_{n_1,n_2} (L_2) &= \frac{\pi^2 \alpha^2}{L_2^2} +\frac{2\pi^2 }{L_2^2} \rl{ n_1^2+n_2^2 +\alpha (n_1-n_2)} \\
P_{n_1,n_2}(B) &= \frac{ e^{-\beta_h E_{n_1,n_2}  (L_1)}} {\sum_{n_1\leq n_2} e^{-\beta_h E_{n_1,n_2}  (L_1)} } \\
P_{n_1,n_2}(A) &= \frac{ e^{-\beta_l E_{n_1,n_2}  (L_2)}} {\sum_{n_1\leq n_2} e^{-\beta_l E_{n_1,n_2}  (L_2)} } \\
\end{aligned}
\end{align}
All the steps mentioned in Section 4, for the case of a single anyon, can be 
repeated in exactly the same manner. The efficiency of the quantum Otto engine
is then 
\begin{align}
\begin{aligned}
\eta_{QOE}=  \frac{W_{out}}{Q_{in}} =1-\frac{\sum_{n_1\leq n_2} E_{n_1,n_2} 
(L_1)(P_{n_1,n_2}(B) -P_{n_1,n_2}(A))  }{ \sum_{m_1\leq m_2} E_{m_1,m_2} 
(L_2) (P_{m_1,m_2}(B) -P_{m_1,m_2}(A))   }
\end{aligned}
\end{align}
Since $ E_{n_1,n_2} (L) \propto \frac{1}{L^2} $, we have 
\begin{align}
\begin{aligned}
E_{n_1,n_2} (L_1) = \frac{L_2^2}{L_1^2 }   E_{n_1,n_2} (L_2) 
\end{aligned}
\end{align}
Therefore the efficiency is 
\begin{align}
\eta_{QOE} & = 1-L_2^2/L_1^2
\end{align}
It is interesting to note that, in this case, the result is essentially the 
the same as the classical result. This is a consequence of the fact that 
energy scales as the inverse square of the length in both cases. 

However, the length $L$ is not the only parameter on which the energy levels
depend. As already mentioned, the strength of the interaction $\alpha$ plays
the same role as the magnetic field in the previous section, and is responsible
for the quantum (anyonic) statistics of the particles. As can be seen from
the expression for the energy spectrum, the dependence of the energy levels 
on $\alpha$ cannot be scaled away. We therefore define a quantum Otto engine in
this case by the following diagram:  
\begin{figure}[h]
\centering
\includegraphics[scale=0.5]{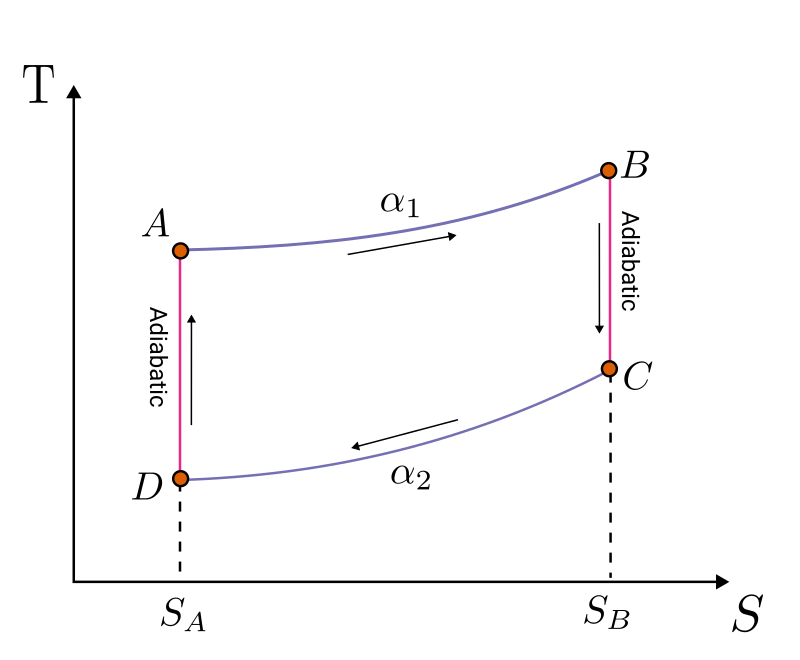}
\caption{The Quantum Otto Engine: fixed volume, variable coupling}
\end{figure}
Once again with all the caveats delineated in the previous examples hold. 
\subsection{Efficiency as a function of the coupling (statistics parameter)
 $ \alpha $}
We are now in a position to compute the efficiency in terms of the statistics
parameter $\alpha$, keeping $L$ fixed. The relevant formulae are
\begin{align}
\begin{aligned}
E_{n_1,n_2} (\alpha_1) &= \frac{\pi^2 \alpha_1^2}{L^2} +\frac{2\pi^2 }{L^2} 
\rl{ n_1^2+n_2^2 +\alpha_1 (n_1-n_2)} \\
E_{n_1,n_2} (\alpha_2) &= \frac{\pi^2 \alpha_1^2}{L^2} +\frac{2\pi^2 }{L^2} 
\rl{ n_1^2+n_2^2 +\alpha_2 (n_1-n_2)} \\
P_{n_1,n_2}(B) &= P_{n_1,n_2}(B) = \frac{ e^{-\beta_h E_{n_1,n_2}  (\alpha_1)}} {\sum_{n_1\leq n_2} e^{-\beta_h E_{n_1,n_2}  (\alpha_1)} } \\
P_{n_1,n_2}(A) &= \frac{ e^{-\beta_l E_{n_1,n_2}  (\alpha_2)}} {\sum_{n_1\leq n_2} e^{-\beta_l E_{n_1,n_2}  (\alpha_2)} } \\
\end{aligned}
\end{align}
The efficiency of the quantum Otto engine can then be written as
\begin{align}
\begin{aligned}
\eta_{QOE}&=  \frac{W_{out}}{Q_{in}} =1-\frac{\sum_{n_1\leq n_2} E_{n_1,n_2} 
(\alpha_1)(P_{n_1,n_2}(B) -P_{n_1,n_2}(A))  }{ \sum_{m_1\leq m_2} E_{m_1,m_2} 
(\alpha_2) (P_{m_1,m_2}(B) -P_{m_1,m_2}(A))   } \\
\end{aligned}
\end{align}
To compute this efficiency, we will need to compute the partition function and 
the sums using theta and partial theta functions -- an exercise we once again
relegate to the Appendix. The result is given by 
\begin{align}
\begin{aligned}
\eta &=  \frac{W_{out}}{Q_{in}} =1-\frac{\sum_{n_1\leq n_2} E_{n_1,n_2} 
(\alpha_1)(P_{n_1,n_2}(B) -P_{n_1,n_2}(A))  }{ \sum_{m_1\leq m_2} E_{m_1,m_2} 
(\alpha_2) (P_{m_1,m_2}(B) -P_{m_1,m_2}(A))   } \\
& =1-\frac{ \frac{\mathcal{X} \rl{\alpha_1,\alpha_1,\beta_h} }
{Z\rl{\alpha_1,\beta_h}} - \frac{\mathcal{X} \rl{\alpha_1,\alpha_2,\beta_l} }
{Z\rl{\alpha_2,\beta_1}} }{ \frac{\mathcal{X} \rl{\alpha_2,\alpha_1,\beta_h} }
{Z\rl{\alpha_1,\beta_h}} - \frac{\mathcal{X} \rl{\alpha_2,\alpha_2,\beta_l} }
{Z\rl{\alpha_2,\beta_1}}   }
\end{aligned}
\end{align}
where
\begin{align}
\begin{aligned}
\mathcal{X} \rl{\alpha,\alpha',\beta} & =  \frac{4\pi^2}{L^2}\rl{4\chi_1\rl{-\frac{\pi^2\beta}{L^2},  0,0} \chi_2 \rl{-\frac{4\pi^2\beta}{L^2},  \alpha/2,\alpha'/2} } \\&+\frac{\pi^2}{L^2}\rl{4\chi_1\rl{-\frac{4\pi^2\beta}{L^2},  -1/2,-1/2} \chi_2 \rl{-\frac{4\pi^2\beta}{L^2},  (\alpha+1)/2,(\alpha'+1)/2} }
\end{aligned}
\end{align}
with
\begin{align}
\begin{aligned}
\chi_1\rl{ \lambda, \gamma, c}=&\sum_{n=-\infty}^{\infty} \rl{ n-c}^2 e^{-\lambda (n-\gamma)^2}   \\=& c^2 e^{-\lambda \gamma^2}\vartheta _3 \rl{e^{2 \lambda \gamma},e^{-\lambda}}+ e^{-\lambda \gamma^2} \frac{(\gamma-c)}{ \lambda}  \frac{\partial }{\partial \gamma} e^{-\lambda \gamma^2}\vartheta _3 \rl{e^{2 \lambda \gamma},e^{-\lambda}} \\&-e^{-\lambda \gamma^2}  \frac{\partial }{\partial \lambda} e^{-\lambda \gamma^2}\vartheta _3 \rl{e^{2 \lambda \gamma},e^{-\lambda}}\\
\end{aligned}
\end{align}
and 
\begin{align}
\begin{aligned}
\chi_2\rl{ \lambda, \gamma, c}=&\sum_{n=0}^{\infty} \rl{ n-c}^2 e^{-\lambda (n-\gamma)^2} \\  =& c^2 e^{-\lambda \gamma^2} \Theta_p\rl{e^{2 \lambda \gamma},e^{-\lambda}}+ e^{-\lambda \gamma^2} \frac{(\gamma-c)}{ \lambda}  \frac{\partial }{\partial \gamma} e^{-\lambda \gamma^2} \Theta_p \rl{e^{2 \lambda \gamma},e^{-\lambda}} \\&-e^{-\lambda \gamma^2}  \frac{\partial }{\partial \lambda} e^{-\lambda \gamma^2}\Theta_p \rl{e^{2 \lambda \gamma},e^{-\lambda}}\\
\end{aligned}
\end{align}
where 
\begin{align}
\begin{aligned}
\vartheta _3 \rl{x,q} &= \sum_{n=-\infty}^{\infty} q^{n^2} x^n \\
\Theta_p(x,q) &=\sum_{n=0}^{\infty} q^{n^2} x^n
\end{aligned}
\end{align}
define the Jacobi theta function, and the partial theta function respectively.

The efficiency is
\begin{align}
\begin{aligned}
\eta_{QOE}&=  \frac{W_{out}O}{Q_{in}} =1-\frac{\sum_{n_1\leq n_2} E_{n_1,n_2} 
(\alpha_1)(P_{n_1,n_2}(B) -P_{n_1,n_2}(A))  }{ \sum_{m_1\leq m_2} E_{m_1,m_2} 
(\alpha_2) (P_{m_1,m_2}(B) -P_{m_1,m_2}(A))   } \\
& =1-\frac{ \frac{\mathcal{X} \rl{\alpha_1,\alpha_1,\beta_h} }
{Z\rl{\alpha_1,\beta_h}} - \frac{\mathcal{X} \rl{\alpha_1,\alpha_2,\beta_l} }
{Z\rl{\alpha_2,\beta_1}} }{ \frac{\mathcal{X} \rl{\alpha_2,\alpha_1,\beta_h} }
{Z\rl{\alpha_1,\beta_h}} - \frac{\mathcal{X} \rl{\alpha_2,\alpha_2,\beta_l} }
{Z\rl{\alpha_2,\beta_1}}   }
\end{aligned}
\end{align}
Since $\alpha_1 = 0$ and $\alpha_2 = 1$, correspond to Bose and Fermi 
statistics respectively, by going through a thermodynamic cycle which changes
the quantum statistics, we specialise to the case of an 
Otto engine based on Bose-Fermi transmutation, as in \cite{koch}. In general,
$\alpha_1$ and $\alpha_2$ can take any real values. 

\section{Conclusions}
In this paper, a detailed study of quantum thermodynamics of small systems
is carried out in the specific context of the quantum Otto engine. The 
working medium is chosen to be one or two anyons in one dimension, whose 
quantum statistics interpolates between the bosonic and fermionic cases. 
Since we accomplish these results using a small number of anyons, we do 
not rely on the macroscopic BEC-BCS crossover studied in \cite{koch}. 

It would be interesting to generalise these results to other thermodynamic
engines. It would also be interesting to choose two-dimensional anyons,
and non-abelian anyons as the working medium. We will report the results of
those cases in the near future. 
\section{Acknowledgements}
This work is partially supported by a grant to CMI from the Infosys
Foundation. 

\appendix

\section{Particle on a Ring Threaded by a Magnetic Field}

We need to compute sums of the form
\begin{align}
\begin{aligned}
Z_j &= \sum_{n=-\infty}^{\infty} e^{ -\beta_j E^j_n} \\
\Upsilon(k,j) &=  \sum_{n=-\infty}^{\infty} E^k_n e^{ -\beta_j E^j_n},~~j,k = h,l\\
\end{aligned}
\end{align}
The Jacobi theta function, defined by
\begin{align}
\begin{aligned}
\vartheta _3 \rl{x,q} &= \sum_{n=-\infty}^{\infty} q^{n^2} x^n
\end{aligned}
\end{align}
may be used to compute the sums. Using this we have
\begin{align}
\begin{aligned}
\sum_{n=-\infty}^{\infty} e^{-\lambda (n-\gamma)^2} =&   e^{-\lambda \gamma^2} \sum_{n=-\infty}^{\infty} e^{-\lambda n^2 +2 \lambda \gamma n}  \\
& =  e^{-\lambda \gamma^2}\vartheta _3 \rl{e^{2 \lambda \gamma},e^{-\lambda}} 
\end{aligned}
\end{align}

\begin{align}
\begin{aligned}
\sum_{n=-\infty}^{\infty} n e^{-\lambda (n-\gamma)^2} & =  e^{-\lambda \gamma^2} \frac{1}{2 \lambda}  \frac{\partial }{\partial \gamma} \sum_{n=-\infty}^{\infty} e^{-\lambda n^2 +2 \lambda \gamma n}  \\
& =  e^{-\lambda \gamma^2} \frac{1}{2 \lambda}  \frac{\partial }{\partial 
\gamma} \vartheta _3 \rl{e^{2 \lambda \gamma},e^{-\lambda}}
\end{aligned}
\end{align}
Also,
\begin{align}
\begin{aligned}
e^{-\lambda \gamma^2}  \frac{\partial }{\partial \lambda} 
\vartheta _3 \rl{e^{2 \lambda \gamma},e^{-\lambda}} & = \sum_{n=-\infty}^{\infty} \rl{ -n^2 + 2 \gamma n} e^{-\lambda (n-\gamma)^2}
\end{aligned}
\end{align}
From this
\begin{align}
\begin{aligned}
 \sum_{n=-\infty}^{\infty} \rl{ n^2} e^{-\lambda (n-\gamma)^2}  & =  
e^{-\lambda \gamma^2} \frac{\gamma}{ \lambda}  \frac{\partial }{\partial 
\gamma} \vartheta _3 \rl{e^{2 \lambda \gamma},e^{-\lambda}} -e^{-\lambda \gamma^2}  \frac{\partial }{\partial \lambda} e^{-\lambda \gamma^2}\vartheta _3 \rl{e^{2 \lambda \gamma},e^{-\lambda}}
\end{aligned}
\end{align}
Therefore
\begin{align}
\begin{aligned}
\sum_{n=-\infty}^{\infty} \rl{ n-c}^2 e^{-\lambda (n-\gamma)^2}   =& 
c^2 e^{-\lambda \gamma^2}\vartheta _3 \rl{e^{2 \lambda \gamma},e^{-\lambda}}+ 
e^{-\lambda \gamma^2} \frac{c\gamma}{ \lambda}  \frac{\partial }
{\partial \gamma} \vartheta _3 \rl{e^{2 \lambda \gamma},e^{-\lambda}} \\&
-e^{-\lambda \gamma^2}  \frac{\partial }{\partial \lambda}\vartheta _3 
\rl{e^{2 \lambda \gamma},e^{-\lambda}}\\
\end{aligned}
\end{align}
The partition function follows immediately:
\begin{align}
\begin{aligned}
Z_j &= \sum_{n=-\infty}^{\infty} e^{ -\beta_j E^j_n}  , ~~j=h,l \\
&=  \sum_{n=-\infty}^{\infty}  e^{ -\beta_j \frac{\hbar^2}{2ma^2} \rl{n- \alpha_j}^2 }  \\
& =  e^{ -\beta_j \frac{\hbar^2}{2ma^2} \alpha_j^2 } \vartheta _3 
\rl{\frac{\alpha_j \beta_j \hbar^2}{2ma^2},e^{-\beta_j \frac{\hbar^2}{2ma^2} }} 
\end{aligned}
\end{align}
with
\begin{align}
\begin{aligned}
\Upsilon(k,j) &=  \sum_{n=-\infty}^{\infty} E^k_n e^{ -\beta_j E^j_n},~~~~
j,k = h,l\\
& =   \sum_{n=-\infty}^{\infty} \frac{\hbar^2}{2ma^2} \rl{n- \alpha_k}^2   
e^{ -\beta_j \frac{\hbar^2}{2ma^2} \rl{n- \alpha_j}^2 } \\
& =  \frac{\hbar^2}{2ma^2} \bigg( c^2 e^{-\lambda \gamma^2}\vartheta _3 
\rl{e^{2 \lambda \gamma},e^{-\lambda}}+ e^{-\lambda \gamma^2} 
\frac{(c\gamma)}{ \lambda}  \frac{\partial }{\partial \gamma} 
\vartheta _3 \rl{e^{2 \lambda \gamma},e^{-\lambda}}\\
& -e^{-\lambda \gamma^2}  \frac{\partial }{\partial \lambda}\vartheta _3 
\rl{e^{2 \lambda \gamma},e^{-\lambda}} \bigg) \bigg|_{c=\alpha_k, \lambda = 
\beta_j \frac{\hbar^2}{2ma^2}, \gamma = \alpha_j }
\end{aligned}
\end{align}
The efficiency is
\begin{align}
\begin{aligned}
  \eta 
  & = 1-\frac{ \frac{\Upsilon (l,h)}{Z_h} -\frac{\Upsilon (l,l)}{Z_l}  }{ \frac{\Upsilon (h,h)}{Z_h} -\frac{\Upsilon (h,l)}{Z_l}  }
\end{aligned}
\end{align}
\section{Two-Anyons on a One-Dimensional Ring}
The partition function is given by
\begin{align}
\begin{aligned}
Z(\alpha,\beta) & = \sum_{n_1\leq n_2} e^{-\beta E_{n_1,n_2}(\alpha)}\\
&= \sum_{n_1\leq n_2} e^{-\beta \rl{  \frac{\pi^2 \alpha^2}{L_1^2} +\frac{2\pi^2 }{L_1^2} \rl{ n_1^2+n_2^2 +\alpha (n_1-n_2)} } }\\
& =  e^{-\beta \rl{  \frac{\pi^2 \alpha^2}{L_1^2} }} \sum_{n_1\leq n_2} e^{-\beta \rl{ \frac{2\pi^2 }{L_1^2} \rl{ n_1^2+n_2^2 +\alpha(n_1-n_2)} } }
\end{aligned}
\end{align}
We define $ m=n_1+n_2 $ and $ n=n_2-n_1 $. We then have
\begin{align}
\begin{aligned}
E_{n_1,n_2}(\alpha) =  E_{m,n}(\alpha_1) = \frac{\pi^2}{L^2} \rl{ n^2 +(m+\alpha)^2} 
\end{aligned}
\end{align}
This gives
\begin{align}
\begin{aligned}
Z(\alpha,\beta)& =  \sum_{ p_1=-\infty}^{\infty} \sum_{ p_2=0}^{\infty} 
e^{ -\beta  \frac{\pi^2}{L^2} \rl{ (2p_1)^2 +(2 p_2+\alpha)^2}  } +  \sum_{ p_1=-\infty}^{\infty} \sum_{ p_2=0}^{\infty} e^{ -\beta  \frac{\pi^2}{L^2} \rl{ (2p_1+1)^2 +(2 p_2+1+\alpha)^2}  } 
\end{aligned}
\end{align}
The first term corresponds to both $ n $ and $ m $ even and the second term corresponds to both $ n $ and $ m $ odd.
The Jacobi theta function and the partial theta function are given by
\begin{align}
\begin{aligned}
\vartheta _3 \rl{x,q} &= \sum_{n=-\infty}^{\infty} q^{n^2} x^n \\
\Theta_p(x,q) &=\sum_{n=0}^{\infty} q^{n^2} x^n
\end{aligned}
\end{align}
Then,
\begin{align}
\begin{aligned}
Z(\alpha,\beta)& =  \sum_{ p_1=-\infty}^{\infty} \sum_{ p_2=0}^{\infty} e^{ -\beta  \frac{\pi^2}{L^2} \rl{ (2p_1)^2 +(2 p_2+\alpha)^2}  } +  \sum_{ p_1=-\infty}^{\infty} \sum_{ p_2=0}^{\infty} e^{ -\beta  \frac{\pi^2}{L^2} \rl{ (2p_1+1)^2 +(2 p_2+1+\alpha)^2}  }  \\
& =   \sum_{ p_1=-\infty}^{\infty} e^{ -\beta  \frac{4\pi^2}{L^2} p_1^2   }  \sum_{ p_2=0}^{\infty} e^{ -\beta  \frac{\pi^2}{L^2} \rl{  4p_2^2 + 4p_2\alpha + \alpha^2}} \\&  +  \sum_{ p_1=-\infty}^{\infty} e^{ -\beta  \frac{\pi^2}{L^2} \rl{  4p_1^2 +4p_1 +1}  } \sum_{ p_2=0}^{\infty} e^{ -\beta  \frac{\pi^2}{L^2} \rl{  4p_2^2 + 4p_2 (\alpha+1) +( \alpha+1)^2}  }
\end{aligned}
\end{align}
can be rewritten in terms of the theta functions as
\begin{align}
\begin{aligned}
Z(\alpha,\beta)& =  e^{ -\beta  \frac{\pi^2 }{L^2} \alpha^2} \vartheta _3 \rl{1,e^{ -\beta  \frac{4\pi^2}{L^2}   } } \Theta_p\rl{ e^{ -\beta  \frac{4\pi^2 \alpha}{L^2}}  ,e^{ -\beta  \frac{4\pi^2}{L^2}} } \\&+ e^{ -\beta  \frac{\pi^2 }{L^2} \rl{(\alpha+1)^2 +1}} \vartheta _3 \rl{e^{ -\beta  \frac{4\pi^2}{L^2}}  ,e^{ -\beta  \frac{4\pi^2}{L^2}   } } \Theta_p\rl{ e^{ -\beta  \frac{4\pi^2 (\alpha+1)}{L^2}}  ,e^{ -\beta  \frac{4\pi^2}{L^2}} }
\end{aligned} 
\end{align}
Let 
\begin{align}
\begin{aligned}
\mathcal{X} \rl{\alpha,\alpha',\beta} & =  \sum_{ p_1=-\infty}^{\infty} \sum_{ p_2=0}^{\infty} \rl{ \frac{\pi^2}{L^2} \rl{ (2p_1)^2 +(2 p_2+\alpha')^2}  } e^{ -\beta  \frac{\pi^2}{L^2} \rl{ (2p_1)^2 +(2 p_2+\alpha)^2}  } \\&+  \sum_{ p_1=-\infty}^{\infty} \sum_{ p_2=0}^{\infty} \rl{ \frac{\pi^2}{L^2} \rl{ (2p_1+1)^2 +(2 p_2+\alpha'+1)^2}  } e^{ -\beta  \frac{\pi^2}{L^2} \rl{ (2p_1+1)^2 +(2 p_2+1+\alpha)^2}  }  \\
\end{aligned}
\end{align}
We have
\begin{align}
\begin{aligned}
\chi_1\rl{ \lambda, \gamma, c}=&\sum_{n=-\infty}^{\infty} \rl{ n-c}^2 e^{-\lambda (n-\gamma)^2}   \\=& c^2 e^{-\lambda \gamma^2}\vartheta _3 \rl{e^{2 \lambda \gamma},e^{-\lambda}}+ e^{-\lambda \gamma^2} \frac{(\gamma-c)}{ \lambda}  \frac{\partial }{\partial \gamma} e^{-\lambda \gamma^2}\vartheta _3 \rl{e^{2 \lambda \gamma},e^{-\lambda}} \\&-e^{-\lambda \gamma^2}  \frac{\partial }{\partial \lambda} e^{-\lambda \gamma^2}\vartheta _3 \rl{e^{2 \lambda \gamma},e^{-\lambda}}\\
\end{aligned}
\end{align}
and 
\begin{align}
\begin{aligned}
\chi_2\rl{ \lambda, \gamma, c}=&\sum_{n=0}^{\infty} \rl{ n-c}^2 e^{-\lambda (n-\gamma)^2} \\  =& c^2 e^{-\lambda \gamma^2} \Theta_p\rl{e^{2 \lambda \gamma},e^{-\lambda}}+ e^{-\lambda \gamma^2} \frac{(\gamma-c)}{ \lambda}  \frac{\partial }{\partial \gamma} e^{-\lambda \gamma^2} \Theta_p \rl{e^{2 \lambda \gamma},e^{-\lambda}} \\&-e^{-\lambda \gamma^2}  \frac{\partial }{\partial \lambda} e^{-\lambda \gamma^2}\Theta_p \rl{e^{2 \lambda \gamma},e^{-\lambda}}\\
\end{aligned}
\end{align}
Therefore
\begin{align}
\begin{aligned}
\mathcal{X} \rl{\alpha,\alpha',\beta} & =  \frac{4\pi^2}{L^2}\rl{4\chi_1\rl{-\frac{\pi^2\beta}{L^2},  0,0} \chi_2 \rl{-\frac{4\pi^2\beta}{L^2},  \alpha/2,\alpha'/2} } \\&+\frac{\pi^2}{L^2}\rl{4\chi_1\rl{-\frac{4\pi^2\beta}{L^2},  -1/2,-1/2} \chi_2 \rl{-\frac{4\pi^2\beta}{L^2},  (\alpha+1)/2,(\alpha'+1)/2} }
\end{aligned}
\end{align}
The efficiency is
\begin{align}
\begin{aligned}
\eta_{QOE}&=  \frac{W_{out}}{Q_{in}} =1-\frac{\sum_{n_1\leq n_2} E_{n_1,n_2} 
(\alpha_1)(P_{n_1,n_2}(B) -P_{n_1,n_2}(A))  }{ \sum_{m_1\leq m_2} E_{m_1,m_2} 
(\alpha_2) (P_{m_1,m_2}(B) -P_{m_1,m_2}(A))   } \\
& =1-\frac{ \frac{\mathcal{X} \rl{\alpha_1,\alpha_1,\beta_h} }
{Z\rl{\alpha_1,\beta_h}} - \frac{\mathcal{X} \rl{\alpha_1,\alpha_2,\beta_l} }
{Z\rl{\alpha_2,\beta_1}} }{ \frac{\mathcal{X} \rl{\alpha_2,\alpha_1,\beta_h} }
{Z\rl{\alpha_1,\beta_h}} - \frac{\mathcal{X} \rl{\alpha_2,\alpha_2,\beta_l} }
{Z\rl{\alpha_2,\beta_1}}   }
\end{aligned}
\end{align}


\end{document}